\documentclass[twocolumn,twoside,slac_two]{revtex4}
\usepackage{graphicx}
\usepackage{fancyhdr}
\begin{document}

\title{Cosmic rays: current status, historical context}

%

\author{Thomas K. Gaisser}
\affiliation{Bartol Research Institute, University of Delaware,\\
Newark, DE 19716, USA}
%

\begin{abstract}
The ISVHECRI conference series emphasizes the connection
between high energy physics and cosmic ray physics---the
study of elementary particles and nuclei from accelerators in the
lab and from space.  In this introductory paper on cosmic rays,
I comment on several current topics in the field
while also providing some historical context.
\end{abstract}

\maketitle

\thispagestyle{fancy}


\section{INTRODUCTION}
The first meeting in the ISVHECRI (International Symposium on Very High Cosmic Ray Interactions) series
was in 1980 in Siberia.  It is my belief, however, that the
Bartol meeting in 1978~\cite{Bartol}
really set the stage for the series by bringing
together some key people from particle physics,
including Carlo Rubbia, David Cline,
J.D. Bjorken, Gaurang Yodh, Larry Jones and Francis Halzen,  and
from cosmic ray physics, including
S. Miyake, C.M.G. Lattes, K. Kamata, Y. Fujimoto, Michael Hillas,
K. Niu and W. Vernon Jones. 
 Many of the same people were together at the second conference
in the series in 1982 which started in La Paz and ended in 
Rio de Janeiro~\cite{ISVHECRI-1982}.
Gaurang Yodh includes a color print of the photo from
the Bartol conference in his Hess Lecture presented at Pune in 2005~\cite{Yodh}, along with
several other historic photos that illustrate his comprehensive review of
cosmic rays.

My charge at this conference is to provide introductory comments on
the status of cosmic rays.  The organization of the paper is conventional, starting with 
inclusive measurements of atmospheric muons and neutrinos, followed by
direct measurements of primary cosmic rays at the top of the atmosphere and
finally extensive air shower measurements in the PeV region and above
where the intensity is too low for direct observations.  Most of these topics 
are covered in detail in papers presented during the conference.  I therefore 
focus on a few points of interest without attempting a comprehensive review.

\section{MUONS AND NEUTRINOS}

An approximate analytic expression for the intensity of atmospheric leptons is
\begin{eqnarray}
&\phi_\iota(E_\iota)& =  \phi_N(E_\iota) \times \nonumber \\
  & & \left\{{A_{\pi\iota}\over 1 + 
B_{\pi\iota}\cos\theta\, E_\iota / \epsilon_\pi}
\,+\,{A_{K\iota}\over 1+B_{K\iota}\cos\theta\, E_\iota / \epsilon_K}\right.\nonumber \\
& & \left. +\,\,\,{A_{{\rm charm}\,\iota}\over 1+B_{{\rm charm}\,\iota}\cos\theta\, 
E_\iota / \epsilon_{\rm charm}}\right\},
\label{angular}
\end{eqnarray}
where $\phi_N(E_\iota) = dN/d\ln(E_\iota)$ is the primary spectrum
of nucleons ($N$) evaluated at the energy of the lepton ($\iota$).  The 
forms are the same for $\iota = \nu$ and $\iota = \mu$ with different
numerical values for the constants $A_{j\iota}$ etc.~\cite{Gaisser,Lipari}.
The three terms correspond respectively to contributions from decay of
pions, kaons and charm~\cite{Desiati}.
The approximations assume that the primary spectrum of nucleons can be
well approximated by a power law, that hadronic cross sections are energy
independent, and that inclusive cross sections for hadro-production
obey Feynman scaling.  The development of this analytical approach dates 
back 50 years to the early papers of Zatsepin~\cite{Zatsepin} and others.
Although detailed simulations are needed in the end, the analytic approximations
provide quantitative insight into the physics of atmospheric leptons.

The classic paper of Barrett {\it et al.}~\cite{Barrett} deals with measurements of muons
in deep detectors that use the overburden to select muons which had high energy
at production in the atmosphere.  They describe how the angular dependence of
the muon intensity deep underground and its dependence on temperature in the stratosphere
depend on the relative contributions of $\pi^\pm$ and $K^\pm$ to the muons.
Both effects arise from the competition between decay and re-interaction
of the parent hadrons in the upper atmosphere.
The expression for the critical energy for meson decay is
\begin{equation}
 E_{\rm critical}\;=\;{\epsilon_j\over \cos{\theta^*}}\,=\,{m_jc^2h_0\over\cos{\theta^*}\,c\tau_j},
\label{critical}
\end{equation}
where $\theta^*$ is the local zenith angle at lepton production taking account
of the curvature of the Earth~\cite{Lipari}.  The parameters 
are $\epsilon_\pi\,\approx 115$~GeV for charged pions as compared to
$\epsilon_K\,\approx 850$ for charged kaons.

\subsection{Temperature Dependence}
In an isothermal approximation, the density  of the atmosphere is described
by an exponential with a scale height of $h_0\approx 6.4$~km, where the numerical
value is applicable to the stratosphere where most high-energy muons and neutrinos
originate.  This numerical value is used to compute the
values of $\epsilon_j$ in Eq.~\ref{critical} that appear in the denominator of
Eq.~\ref{angular}.
 From the ideal gas equation relating density and pressure, one finds
$h_0\;=\;R\,T$.   Since $\epsilon\,\propto\,h_0$, this
relation gives rise to a seasonal variation in
the rate of muons in a deep underground detector through the dependence
of $\epsilon$ in Eq.~\ref{angular} on $h_0$.  Rates can also reflect
sudden changes in the upper atmosphere~\cite{sudden}.  

The magnitude of the correlation 
with temperature depends on the muon energy at production and hence on the depth of the
detector.  Because $\epsilon_K\,> \epsilon_\pi$, the kaon contribution
has a weaker correlation with temperature than the pion contribution,
which means that the magnitude of the correlation with temperature depends
on the kaon to pion ratio.  This dependence has been discussed recently
by the MINOS Collaboration in connection with their measurements of cosmic-ray 
muons~\cite{MINOS} in the far detector at Soudan, where the minimum muon
energy to reach the detector is approximately 1 TeV.
In IceCube the seasonal variation in the muon rate is approximately $\pm$8\%~\cite{Tilav}.
This is an important feature of the background that must be accounted for
in interpreting the data.

Rates of atmospheric neutrinos are also expected to vary with temperature~\cite{Bernardini}.
IceCube is large enough so that the
rates of TeV neutrinos may be sufficiently high to actually measure their 
seasonal variation~\cite{Desiati}.  
Since the neutrinos are detected primarily as upward moving
$\nu_\mu$-induced muons (using the earth
as a filter against the downward muon background), the magnitude and phase of
their variation will depend on the part of the sky from which they originate.

Because the lifetime of charmed hadrons is so short, their decay products are ``prompt."
The denominator of the charm contribution in Eq.~\ref{angular} is unity for
$E_\iota < 10$~PeV.  As a consequence, the prompt muons and neutrinos are isotropic
and have a harder spectrum than decay products of pions and kaons in the
energy range above a TeV.  These two features have long been used to search for
a prompt contribution to atmospheric muons.  In Ref.~\cite{Desiati} it is pointed out
that the lack of seasonal variation is also a signature of the charm channel.

\subsection{Muon Charge Ratio}

Both MINOS~\cite{MINOSpm} and OPERA~\cite{OPERA} are sufficiently deep so 
that they measure the charge ratio of muons that had energies in the TeV
range at production in the atmosphere.  The measured increase in the
charge ratio from 1.27 below 100 GeV to 1.37 above a TeV
reflects the enhanced importance of kaons at high energy.
In particular, associated production,
\begin{equation}
 p\,+\,air\,\rightarrow\,\Lambda\, +\, K^+\, +\, X,
\label{associated}
\end{equation}
is an important factor in making
 the $+/-$ charge ratio larger for kaons than for pions.
A measurement of the charge ratio by CMS at somewhat lower energy
was reported at this conference~\cite{CMS}.  The results from MINOS
on temperature dependence as well as charge ratio were summarized here
by Schreiner~\cite{Schreiner}.

The strong asymmetry of $K^+$ compared to $K^-$ in the forward
fragmentation region (which dominates production of secondaries
from a steep spectrum) reflects the fact that the $\Lambda$ 
has constituents in common with the incident proton, while there
is no corresponding channel for production of $K^-$.  The
charmed analog of Eq.~\ref{associated} is also highly asymmetric~\cite{SELEX}.
The possible consequence for prompt atmospheric leptons
is a subject of current interest.

Analytic expressions for charge separated intensities ($\mu^+,\;\mu^-$ and 
$\nu_\mu,\;\bar{\nu}_\mu$) can be constructed in parallel with Eq.~\ref{angular},
as was worked out for pions in Ref.~\cite{Frazer}.  The charge-sign dependence
starts with the separation of neutrons from protons in the primary cosmic-ray
beam and tracks the charge dependences of meson production and decay~\cite{Lipari}.

The new data on temperature dependence and charge sign dependence
of atmospheric muons call for a global fit to all the relevant parameters.
These include all
the spectrum-weighted moments, $Z_{j\iota}$, the attenuation lengths
and also the parameters that
reflect the initial charge-sign dependence imparted by the fraction of primary
nuclei with $Z>1$ and how this changes with energy.

\section{GALACTIC COSMIC RAYS}

\subsection{Direct measurements of the primary spectrum}

Measurements by ATIC~\cite{ATIC} and CREAM~\cite{CREAM} presented by
Seo at this meeting~\cite{Seo} are providing a new and more complex picture of
primary composition in the multi-TeV energy region than was generally
assumed in the past.  In particular, spectra seem to harden somewhat
above 200 GeV/nucleon, and there is a hint that the spectrum of helium is somewhat
harder than that of protons between 10 and 100 TeV~\cite{CREAM}.

\begin{figure}
\includegraphics[width=65mm]{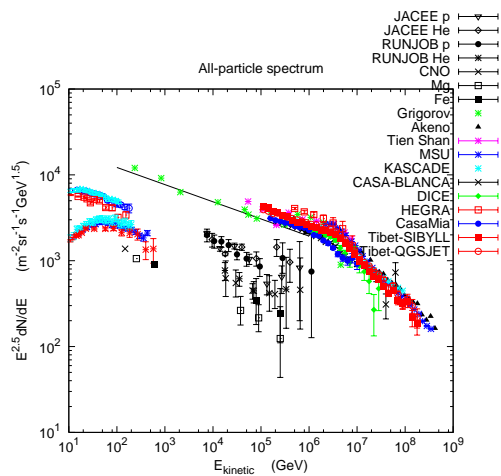}
\caption{Summary of measurements before 2005.}
\label{allparticle2T}
\end{figure}

Figure~\ref{allparticle2T} shows the situation a few years ago.  Up to 100 GeV
there are rather precise spectrometer measurements of protons
and helium from BESS~\cite{BESS}, AMS~\cite{AMS}, CAPRICE~\cite{CAPRICE} and
others.  The measurements of BESS and AMS agreed well with each other in normalization as
well as spectrum, with proton intensities somewhat higher than CAPRICE and earlier
measurements.  The line in Fig.~\ref{allparticle2T} is an extrapolation to high energy
of the sum of the individual spectra assuming that all are described by a power
law with integral spectral index $\gamma=1.7$.  Normalization of protons and helium
are to BESS and AMS.  Normalization of other nuclei are obtained from HEAO~\cite{HEAO}
and other measurements as summarized in~\cite{PDG}.  The extrapolation comes in
on the low side of the indirect measurements with air shower experiments around
a PeV, just below the knee.  

The plot also shows the early measurements of
JACEE and RUNJOB in the region from 10 to several hundred TeV.
There was a significant discrepancy between JACEE and RUNJOB for 
the flux of helium.  ATIC-2 and CREAM seem to confirm the higher
He intensity of JACEE.  Adding together the CREAM measurements of
protons, helium and heavier nuclei gives a higher all-particle intensity
around 100 TeV in better agreement with the TIBET air shower measurement.
The harder spectrum reported by CREAM in the range 10 - 100 TeV/nucleus when
extrapolated to the knee region leads to this improved agreement.

\subsection{Knee of the spectrum}

Fifty years ago Peters~\cite{Peters} suggested that the steepening in the knee region
may be associated with the upper end of the 
spectrum of particles from the dominant source of
cosmic rays, which could be powered by supernovas.
He explained the relatively smooth behavior of
the cutoff as a consequence of different primary nuclei
having cutoffs at different energy per particle
under the assumption that the underlying effect depends
on magnetic rigidity.  The definition of rigidity, 
$R(Z) = Pc/Ze$, provides the scaling relation between total energy
($E\approx Pc$) and the charge $Ze$ of nuclei moving in
a given configuration of magnetic fields.
If the ``cutoff" of the spectrum occurs at a characteristic
rigidity, $R_c$, then protons bend first at $E_c=eR_c$
followed by helium at $2eR_c$, carbon at $6eR_c$ up to
iron at $26eR_c$.  This sequence is sometimes referred
to as a ``Peters cycle".  The KASCADE Experiment~\cite{kascade} provided
the first observation of this sequence of changes in
the knee region, with protons steepening first then helium and
then heavier nuclei.  The relative abundances of various nuclei
inferred from the KASCADE data depends on the interaction model
used for simulation necessary to unfold the spectra.  However,
helium appears more abundant than protons above 1 PeV
whether SIBYLL or QGSjet is used for the unfolding.
This feature is qualitatively consistent with an extrapolation
of the direct measurements assuming a harder spectrum for helium
than for protons.

\begin{figure}[thb]
\includegraphics[width=65mm]{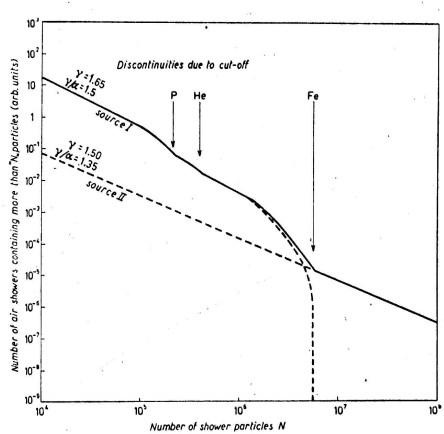}
\caption{Two component concept of Peters (1960).}
\label{Fig2}
\end{figure}

Already at the time of Peters' paper 
the spectrum was known to continue beyond the knee. 
He suggested a second population with a significantly smaller
total number of particles but with a slightly harder spectrum
that continues to higher energy,
as shown in Fig.~\ref{Fig2} from his paper.  We now know, however,
that the spectrum continues for three orders of magnitude after
the knee with a {\em steeper} spectrum than below the knee.
Not until the ankle does the spectrum harden in the way
expected by Peters.
At least one additional Peters cycle with a cutoff at higher
energy is needed to fill in the gap between the knee and the ankle.
Hillas~\cite{Hillas} calls this ``Component B".

\subsection{Component B}

\begin{figure}
\includegraphics[width=75mm]{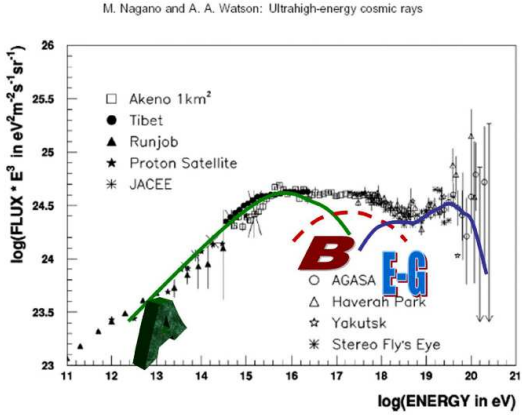}
\caption{Schematic picture showing three populations of
cosmic rays: ${\cal A}$ is the dominant galactic component,
perhaps accelerated directly by supernova remnants; ${\cal B}$
is a higher-energy galactic population of uncertain origin;
${\cal E-G}$ is the extra-galactic component that dominates
above the ankle around 3 EeV.}
\label{Fig3}
\end{figure}

The question of the cosmic-ray spectrum above the knee was
extensively discussed at the two meetings on ``Physics
at the end of the Galactic cosmic ray spectrum" in Aspen 
in 2005 and 2007.  Figure~\ref{Fig3} illustrates the discussion
with schematic outlines of three separate components
superimposed on the plot of the spectrum from the review of
Nagano and Watson~\cite{NW}.  In this scheme, the second
knee is a cutoff associated with a second
population of Galactic sources of unknown origin, and
the extra-galactic
component dominates only above the ankle.
It is interesting that the new spectrum of KASCADE-Grande
presented at this conference~\cite{Arteaga} shows a 
concavity just above 10 PeV that could be interpreted
as the onset of Component B.

Another point is to estimate the power required to produce the 
B-component.  The desired result is obtained by
integrating the spectrum of the B component weighted by
an estimate of the lifetime of high-energy cosmic rays
in the galaxy.  The expression
\begin{equation}
{\cal L} \;=\;V_G\,{4\pi\over c}\int E{1\over\tau(E)}{{\rm d}N\over{\rm d}E}{\rm d}E
\label{power}
\end{equation}
gives the total power in the sources to supply this component.
Taking
\begin{eqnarray}
\tau_{esc}= 2\times 10^7\,{\rm yrs}\,E^{-0.33}, \\
V_{Galaxy} \,=\,3\times 10^{66}\,{\rm cm}^3
\end{eqnarray}
and a differential source spectral index below
the cutoff of $\gamma + 1 = -2.1$, 
leads to a 
power requirement of $\sim 2\times 10^{39}$~erg/s.

\subsection{Acceleration mechanisms}
Another topic featured at the Aspen meetings was a discussion
of acceleration mechanisms.  The classic estimate of the
upper limiting energy of acceleration by shocks driven
into the interstellar medium by expanding supernova remnants
is $E_{\rm max} \approx 100\,{\rm TeV}$ for protons~\cite{LagageCesarsky}.
Two aspects of non-linear shock acceleration are important
for building a synoptic picture of Galactic cosmic rays.
One is the amplification of magnetic fields near the shock~\cite{Bell},
which increases $E_{\rm max}\propto ZeBR$.  This allows diffusive
acceleration driven by supernova remnants to accelerate particles
to a maximum rigidity one PV or higher, as would be expected if the
knee is related to the upper limit of the dominant population of
Galactic cosmic rays.

The other important effect is 
the concavity introduced by the cosmic-ray pressure of the
highest energy particles on the plasma upstream of the shock.
The result is that higher energy particles experience a
larger velocity difference than the canonical factor of 4
at the sub-shock.  In diffusive shock acceleration, the
integral spectral index $\gamma$ is related to the ratio
of velocity of the upstream flow into the acceleration
region ($u_1$) to the downstream velocity ($u_2$) by
\begin{equation}
\gamma\;=\;{3\over u_1/u_2\,-1}.
\label{gamma}
\end{equation}
In the non-linear case, $u_1(d)$ is a function of
distance upstream of the shock, with $u_1(d)/u_2$
increasing as $d$ increases.  Higher energy particles
can diffuse further upstream on average before reversing
direction; hence the concave nature of the spectrum~\cite{Blasi}. 

Non-linear diffusive shock acceleration is contrasted with
the test-particle approximation in which the effect of the
accelerated particles on the plasma flowing through the
shock is neglected.  Then, for a strong shock, $u_1/u_2=4$
and $\gamma=1$, which produces a spectrum with equal
energy content per logarithmic interval on energy.  In the
non-linear case, most of the energy of the accelerated
spectrum is near $E_{\rm max}$.  An interesting idea that uses
this feature is the proposal by Ptuskin \& Zirakashvili~\cite{PZ}.
At each epoch in the evolution of a supernova remnant, most of
the power goes into particles at the highest possible energy.
As the remnant evolves, $E_{\rm max}$ decreases and the highest
energy particles can then escape upstream (that is, to the outside).
The spectrum produced by a supernova over its lifetime is then
the sum of the contribution from each epoch.

A full picture of the sources, acceleration and propagation
of galactic cosmic rays does not yet exist.  What seems likely
to me, however, is that in reality there are many sources with different
spectra and capable of accelerating particles to different
maximum energies.  The complex structure of the spectrum indicated by the CREAM
results below 100 TeV and by the KASCADE-Grande results in the 
air shower range may reflect such underlying complexity.

\section{EXTRAGALACTIC COSMIC RAYS}

A well-known alternative to the picture described in the previous 
section and summarized in Fig.~\ref{Fig3} is that of Berezinsky
and collaborators~\cite{Berezinsky}.  They explain the ankle as a consequence of
energy loss by protons due to $e^\pm$ pair production on the
cosmic background radiation.  Such a model requires that the 
extra-galactic component consists primarily of protons and that
there is a cosmological distribution of sources capable of
accelerating particles to energies above $100$~EeV.
In this picture the extra-galactic population dominates down
to $\sim 300$~PeV and a galactic B component is not needed.
A signature would be a transition in the composition from heavy nuclei at the
end of the galactic population to protons that dominate the extra-galactic
component in this model.

\subsection{Composition}

Results from HiRes~\cite{HiRes} and HiRes-MIA~\cite{MIA} do indicate
a transition from heavy nuclei below $300$~PeV to protons in the EeV
range and above.  But the question is unsettled because Auger results~\cite{Auger}
indicate on the contrary that the composition above $10$~EeV is mostly heavy nuclei.

The two contrasting pictures of the transition from galactic to extra-galactic
cosmic rays and their relation to composition have been discussed in a series
of papers by Allard {\it et al.}~\cite{Allard}.  In the case of a mixed composition at the
source, even if there is a preponderance of heavy nuclei,
 a significant fraction of protons is expected at Earth as a result
of photo-nuclear disintegration during propagation through the microwave background,
provided that 
the sources have a cosmological distribution and accelerate particles to
sufficiently high energy.  In any case, this picture would also be difficult to reconcile
with the Auger results on composition because there would be large fluctuations
in the Xmax distribution as a consequence of the mixed composition after 
propagation to Earth (protons and iron).

There is considerable experimental activity to explore the transition region.
Low-energy extensions of Auger are now in operation~\cite{HEAT-AMIGA}, and the low-energy extension
of the telescope array is under construction~\cite{TALE}.  Meanwhile,
KASCADE-Grande~\cite{Arteaga}, IceCube~\cite{TilavI3} 
and TUNKA~\cite{TUNKA} are reaching the transition region from below.
Concerning the discrepancy in composition between HiRes and Auger,
it may be relevant to note that the selection of events is somewhat different
in the two cases.  HiRes reconstructs event trajectories using two Fly's Eyes
in stereo, while the Auger reconstruction uses a hybrid technique that
involves times of hits in one or more surface detectors along with the fluorescence
signal.  Some insight on the differences may come from the Telescope
Array~\cite{TA}, which is also a hybrid detector.

\begin{figure}
\includegraphics[width=75mm]{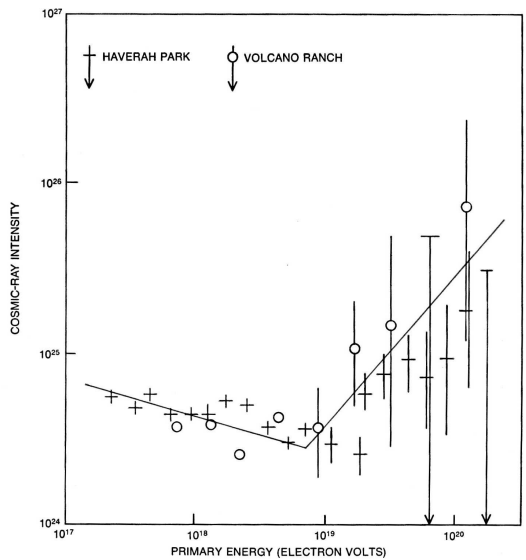}
\caption{Summary of primary spectrum measurements from Linsley's review~\cite{LinsleyReview}.}
\label{Fig4}
\end{figure}

\subsection{Two generic models}

One possibility is a machine in which the magnetic fields required for
acceleration are strong enough and in an appropriate configuration so that
accelerated protons and nuclei lose energy in the ambient radiation
fields by photo production and photo disintegration respectively
before they escape from the system.  The acceleration time scale has
to be short or comparable to the photo-interaction rate.  Then if
the escape time for neutrons is short compared to their photo-interaction
time scale they will escape and contribute to the population of
cosmic rays.  In this class of models the extra-galactic cosmic
rays would be protons from decay of the escaping neutrons.

In addition, neutrinos produced in the same sources in reaction chains like
\begin{equation}
p\,+\,\gamma\;\rightarrow\;n\,+\,\pi^+\;\rightarrow\;n\,+\mu^+\,+\nu_\mu
\label{p-nu}
\end{equation}
will then have a spectrum and intensity related by kinematics to
that of the cosmic-rays.  This class of models would be a realization
of the Waxman-Bahcall limit~\cite{WB}.  Ahlers {\it et al.}~\cite{Ahlers}
point out that the limits from IceCube already constrain models of
this type.

A contrasting possibility is that the extra-galactic cosmic rays 
could be powered 
by shocks driven into an external environment.  In this case, whatever
material is available would be accelerated, in analogy with galactic
cosmic rays accelerated in shocks driven by supernova explosions.
External shocks driven by gamma-ray bursts or jets of active galaxies~\cite{Berezhko}
could be realizations of this class of models.  The composition
could be mixed and there would be no particular reason to expect
a high intensity of neutrinos or gamma-rays from interactions
of the accelerated particles.

\begin{figure}
\includegraphics[width=75mm]{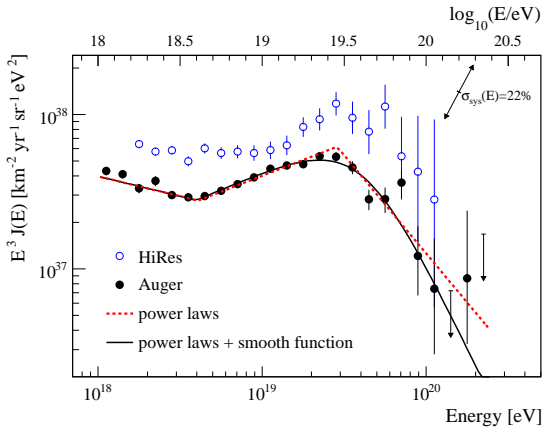}
\caption{HiRes and Auger spectrum measurements from Ref.~\cite{AugerSpectrum}.}
\label{Fig5}
\end{figure}

\subsection{The end of the cosmic-ray spectrum}

The experimental discovery of extra-galactic cosmic rays
can be attributed to the Volcano Ranch experiment, built
by John Linsley and Livio Scarsi starting in 1958~\cite{archive}.  The array
was spread over an area of approximately 8 km$^2$.  In 1962 they
detected an event estimated to have an energy of
$10^{20}$~eV (with large uncertainty), too high
to be contained in the galaxy~\cite{LinsleyPRL}.  
In his 1978 review of ultra-high energy cosmic rays,
Linsley~\cite{LinsleyReview} combined data from Volcano Ranch and Haverah
Park~\cite{HP} in the U.K. into a plot that shows the ankle of the spectrum,
which he interpreted as the onset of extragalactic cosmic rays.
Linsley's figure is reproduced here
as Figure~\ref{Fig4}.  The data clearly show the ankle 
but no sign of the suppression expected from 
photo-production interactions in the microwave background if the
highest energy particles come from a cosmological distribution
of sources.

Linsley's review came at an important time in the history of
measurements of ultra-high-energy cosmic rays.  The
Fly's Eye detector was under construction at
Dugway Proving Ground in Utah after the technique had been demonstrated for the
first time by using three new Fly's Eye receivers at the
Volcano Ranch array to see the fluorescence light from
air showers in coincidence with Linsley's ground array.~\cite{FEtest}.

It was another thirty years until HiRes, using the Fly's Eye technique,
showed a drop in the spectrum~\cite{HiResSpectrum} consistent with that predicted by
Greisen~\cite{Greisen} and Zatsepin \& Kuz'min~\cite{ZK}.  The HiRes
result is shown in Fig.~\ref{Fig5} along with the latest spectrum
from the Auger detector~\cite{AugerSpectrum}.  The two spectra agree well with
each other within the uncertainties of the energy calibration, which
are of the order of 20\%.  
Moving one of the sets of points along the direction of the slanted arrow
brings the two measurements into agreement.
In fact this point illustrates the possibility noted already by
Linsley of using the GZK feature to calibrate the energy assignment.

\subsection{Anisotropy?}

Linsley also pointed out that protons with energies of the order of 100 EeV
would have sufficiently high rigidity to be likely to point back
to their sources within $\sim100$~Mpc.  In this connection it is 
interesting to compare the sky plot in Linsley's review with that of
the Auger experiment.  Linsley shows the directions of 16 events
with energy greater than 50 EeV measured by Haverah Park~\cite{HP-directions}.
He points out that a significant fraction of the events come in
from directions at large angle to the galactic plane, which argues in
favor of an extragalactic origin if the particles are protons with
high rigidity.  The first Auger sky plot published in 2007~\cite{Auger1} 
showed a rather strong correlation of events with $E\ge57$~EeV
with the large scale distribution of matter in the nearby universe
as traced by AGNs.  Since then the significance of the correlation
has decreased~\cite{Auger2}, but it is still suggestive of a correlation
of the kind that might be expected if the sources are correlated with
the large-scale distribution of matter~\cite{Stanev} and if the particles in this
energy range are mostly protons.  

In this connection, it is important
to note that the measurement showing heavy composition extends only up to
40~EeV and does not yet overlap with the event sample used to look
for anisotropy.   In a recent paper, Calvez, Kusenko and Nagataki~\cite{Kusenko}
suggest that GRBs in the Milky Way on a $10^5$ year time scale could contribute
to the pool of cosmic rays above 10 EeV.  This would be a kind of second ``B component"
that would explain the heavy composition of Auger as the end of the highest
energy galactic Peters cycle.  The onset of the true extragalactic component would
then be around 50 EeV.  The challenge is to find ways to increase the statistics
and precision of measurements of cosmic rays up to 100 EeV despite the
extremely low intensity of such particles.

\subsection{End of the spectrum revisited}

Finally, it is salutary to remember the Hillas plot~\cite{HillasPlot}.
In his iconic diagram, Hillas places potential sources in a space defined
by their sizes and magnetic fields.  A remarkable feature of the
plot is that the few sources capable of accelerating particles to 100~EeV
are not far above the line.  A realistic possibility is, therefore, that
the steepening of the spectrum may to some extent reflect a cutoff at the accelerators
as well as (or instead of) the effect of energy losses to interactions
with photons of the cosmic microwave background.

\vspace{-.5cm}
\begin{acknowledgments}
I am grateful to Adrienne Kolb and Eun-Joo Ahn for an enlightening visit
to the Fermilab archive.  I thank Larry Jones for information
about the ISVHECRI conference series and Ralph Engel for bringing Reference~\cite{Kusenko}
to my attention. I thank  Todor Stanev for reading this paper in draft form and
Hank Glass for generous editorial assistance with this paper.  Much of the material 
for this paper was developed during my stay in Germany with financial support 
from the Humboldt Foundation.  My work is also
 supported in part by the U.S. Department of Energy 
under contract DE-FG02-91ER40626.
\end{acknowledgments}

\bigskip 

\end{document}